\documentclass[aps,prl,reprint,nopacs,amsmath,amssymb,superscriptaddress]{revtex4-1}%

\usepackage{multirow}
\usepackage{bigstrut}
\usepackage{amssymb}
\usepackage{amsmath}
\usepackage{graphicx}
\usepackage{dcolumn}
\usepackage{bm}
\usepackage{CJK}
\usepackage{relsize}
\usepackage{cancel}
\usepackage{color}
\usepackage{setspace}
\usepackage{times}
\usepackage{mathptmx}

\begin{document}

\title{Modulation of pure spin currents with a ferromagnetic insulator}

\author{Estitxu Villamor}\affiliation{CIC nanoGUNE, 20018 Donostia-San Sebastian, Basque Country, Spain}
\author{Miren Isasa}\affiliation{CIC nanoGUNE, 20018 Donostia-San Sebastian, Basque Country, Spain}
\author{Sa\"ul V\'elez}\affiliation{CIC nanoGUNE, 20018 Donostia-San Sebastian, Basque Country, Spain}
\author{Amilcar Bedoya-Pinto}\affiliation{CIC nanoGUNE, 20018 Donostia-San Sebastian, Basque Country, Spain}
\author{Paolo Vavassori}\affiliation{CIC nanoGUNE, 20018 Donostia-San Sebastian, Basque Country, Spain}\affiliation{IKERBASQUE, Basque Foundation of Science, 48011 Bilbao, Basque Country, Spain}
\author{Luis E. Hueso}\affiliation{CIC nanoGUNE, 20018 Donostia-San Sebastian, Basque Country, Spain}\affiliation{IKERBASQUE, Basque Foundation of Science, 48011 Bilbao, Basque Country, Spain}
\author{F. Sebasti\'an Bergeret}\affiliation{Centro de F\'isica de Materiales (CFM-MPC) Centro Mixto CSIC-UPV/EHU, 20018 Donostia-San Sebastian, Basque Country, Spain}\affiliation{Donostia International Physics Center (DIPC), 20018 Donostia-San Sebastian, Basque Country, Spain}
\author{F\`{e}lix Casanova}\affiliation{CIC nanoGUNE, 20018 Donostia-San Sebastian, Basque Country, Spain}\affiliation{IKERBASQUE, Basque Foundation of Science, 48011 Bilbao, Basque Country, Spain}

\begin{abstract}
\small{We propose and demonstrate spin manipulation by magnetically controlled modulation of pure spin currents in cobalt/copper lateral spin valves, fabricated on top of the magnetic insulator Y$_3$Fe$_5$O$_{12}$ (YIG). The direction of the YIG magnetization can be controlled by a small magnetic field. We observe a clear modulation of the non-local resistance as a function of the orientation of the YIG magnetization with respect to the polarization of the spin current. Such a modulation can only be explained by assuming a finite spin-mixing conductance at the Cu/YIG interface, as it follows from the solution of the spin-diffusion equation. These results open a new path towards the development of spin logics.}
\end{abstract}

\pacs{}
\maketitle

\emph{Spintronics} is a rapidly growing field that aims at using and manipulating not only the charge, but also the spin of the electron, which could lead to faster data processing speed, non-volatility and lower electrical power consumption as compared to conventional electronics \cite{1}. Sophisticated applications such as hard-disk read heads and magnetic random access memory (MRAM) have been introduced in the last two decades. 

Further progress could be achieved with pure spin currents, which are an essential ingredient in an envisioned spin-only circuit that would integrate logics and memory \cite{2}. The most basic unit in such a concept is the spin analog to the transistor, in which the manipulation of pure spin currents is crucial. The original proposal by Datta and Das \cite{3}, which is also applicable to pure spin currents \cite{4}, suggested a spin manipulation that would arise from the spin precession due to the spin-orbit interaction modulated by an electric field (Rashba coupling). However, a fundamental limitation appears here, because the best materials for spin transport are those showing the lowest spin-orbit interaction and, therefore, there has been no success in electrically manipulating the spins and propagating them at the same environment, with few exceptions \cite{4}.

Alternative ways to control pure spin currents are thus desirable. One could take advantage of the spin-mixing conductance concept \cite{5,6} at nonmagnetic metal (NM)/ferromagnetic insulator (FMI) interfaces, which governs the interaction between the spin currents present at the NM and the magnetization of the FMI. This concept is at the basis of new spin-dependent phenomena, including spin pumping \cite{6,7,8,9,10,11,12}, spin Seebeck effect \cite{6,13}, and spin Hall magnetoresistance (SMR) \cite{6,14,15,16,17,18}. In these cases, a NM with large spin-orbit coupling is required to convert the involved spin currents into charge currents via the inverse Spin Hall effect \cite{19}. 

\begin{figure}[t]
\centering
\includegraphics[scale=1]{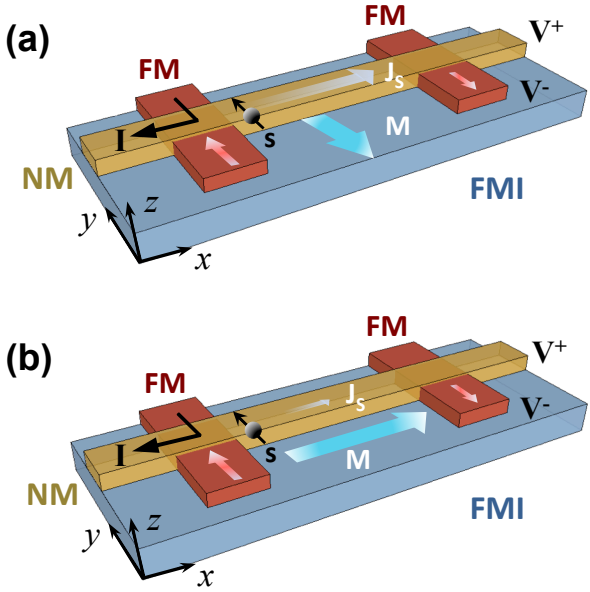}
\caption{\small{(Color online) Scheme of the device used to modulate a pure spin current with magnetic gating. It consists of a ferromagnetic (FM)/ nonmagnetic (NM) lateral spin valve on top of a ferromagnetic insulator (FMI). The nonlocal measurement configuration is shown. The $x$, $y$ and $z$ axes are indicated as used in the text. (a) When the magnetization of the FMI ($M$) and the polarization ($s$) of the injected pure spin current ($j_s$) are parallel, there will be no spin absorption. (b) When $M$ and $s$ are perpendicular, the spin absorption will be maximum. }}\label{1}
\end{figure}

In this Rapid Communication, we demonstrate an alternative way of modulating pure spin currents based on magnetic, instead of electric, gating. To that end, we use lateral spin valves (LSVs). These devices allow an electrical injection and detection of pure spin currents in a NM channel by using ferromagnetic (FM) electrodes in a nonlocal configuration \cite{20,21,22,23,24,25,26,27,28,29}. The LSVs have been fabricated on top of a FMI, in order to enable the magnetic gating of the pure spin currents. The basic idea is depicted in Fig. \ref{1}: when the spin polarization ($s$) has the same direction as the magnetization ($M$) of the FMI, the spin current reaching the detector will not vary with respect to the case where no FMI is used [Fig. \ref{1}(a)]. However, when $s$ and $M$ are noncollinear, part of the spin current will be absorbed by $M$ via spin-transfer torque \cite{30,31,32}, leading to maximum spin absorption for perpemdicular $M$ and $s$ [Fig. \ref{1}(b)]. By using LSVs, we are able to extract the spin-mixing conductance of NM/FMI interfaces in the absence of charge currents, which otherwise could lead to spurious effects, as suggested by some authors \cite{33,34}. Furthermore, the use of NM metals with low atomic number, employed in LSVs, rules out spin-orbit interaction effects that might exist for other systems, such as Pt/YIG \cite{35}.

\begin{figure}[t]
\centering
\includegraphics[scale=1]{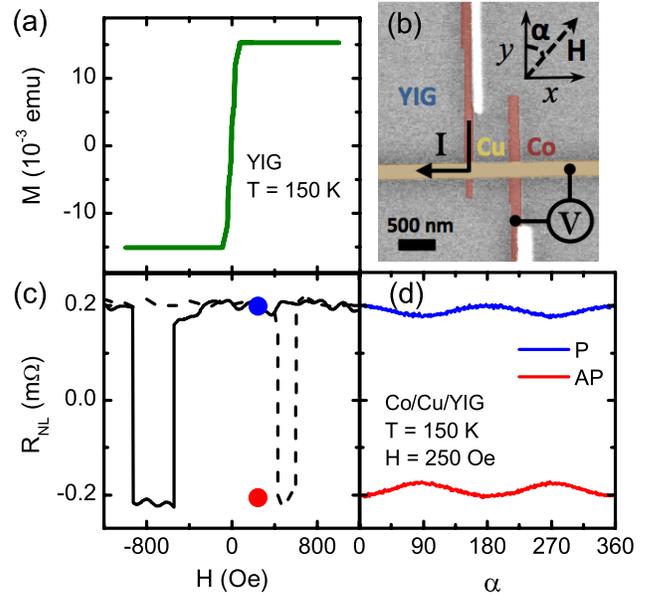}
\caption{\small{(Color online) (a) Magnetization of YIG ($M$) as a function of the applied in-plane magnetic field $H$ measured at 150 K. (b) Colored scanning electron microscopy (SEM) image of a LSV. The nonlocal measurement configuration, materials, direction of $H$ and its angle $\alpha$ with respect to the FM electrodes are shown. (c) Nonlocal resistance ($R_{NL}$) measured at 150 K as a function of $H$ with $\alpha=0^{\circ}$ for a LSV with a separation distance between Co electrodes of $L=1.6 \text{ }\mu \text{m}$. The solid (dashed) line indicates the decreasing (increasing) sweep of $H$. A constant background of $0.14\text{ m} \Omega$ is subtracted from the data. Blue and red dots correspond to the value of $R_{NL}$ at the parallel (P) and antiparallel (AP) configurations of the Co electrodes, respectively, at $H=250$ Oe. (d) $R_{NL}$ as a function of $\alpha$, measured for both the P and AP configurations, at 150 K with $H=250$ Oe for the same LSV.}}\label{2}
\end{figure}

We chose Y$_3$Fe$_5$O$_{12}$ (YIG) \cite{36} as a magnetic gate because it is ferromagnetically soft and has a negligible magnetic anisotropy. $M$ as a function of the applied in-plane magnetic field ($H$) measured by a vibrating sample magnetometer (VSM) saturates at $\sim100$ Oe [Fig. \ref{2}(a)], allowing control of $M$ above this field. Cobalt (Co)/copper (Cu) LSVs were fabricated on top of YIG by two-step electron-beam lithography, ultrahigh-vacuum evaporation, and a lift-off process [Fig. \ref{2}(b)] \cite{37}. Ar-ion milling was performed prior to the Cu deposition in order to remove resist leftovers \cite{37}. To overcome the low spin injection of Co when using transparent interfaces \cite{21,22,23}, an oxide layer was created at the Co/Cu interface by letting Co oxidize after milling and before Cu deposition. The presence of an interface resistance, estimated to be $R_I\geq5\text{ }\Omega$, is known to enhance the spin injection efficiency \cite{24,25}. The LSVs were bridged by the same Cu channel, with thickness $t\sim100$ nm, width $w\sim200$ nm, and different edge-to-edge distances ($L$) between the FM electrodes \cite{37}. 

All measurements were performed using a "dc reversal" technique \cite{27} in a liquid-He cryostat with an applied magnetic field $H$ at a temperature of 150 K. The sample can be rotated in plane in order to change the direction of $H$, which is given by the angle $\alpha$ defined in Fig. \ref{2}(b). The nonlocal voltage $V_{NL}$ measured at the detector, normalized to the injected current $I$, is defined as the nonlocal resistance $R_{NL}= V_{NL}/I$ [Fig. \ref{2}(b) shows a measurement scheme]. First, in order to check the standard performance of the LSV, the direction of $H$ was fixed parallel to the FM electrodes ($\alpha=0^{\circ}$) and its value was swept from positive to negative, and vice versa, while $R_{NL}$ was measured. This is plotted in Fig. \ref{2}(c), where $R_{NL}$ changes from positive to negative when the relative magnetization of the FM electrodes changes from parallel (P) to antiparallel (AP) by sweeping $H$. This measurement is an unambiguous demonstration that a pure spin current is transported along the Cu channel \cite{20,21,22,23,24,25,26,27,28,29}. It is worth noting that the relative magnetization of the Co electrodes changes at $H\geq400$ Oe, far above the saturation field of YIG ($\sim100$ Oe). This detail is important for the performance of the next measurement, which consists in measuring $R_{NL}$ while fixing the value of $H$ and sweeping $\alpha$. As shown in Fig. \ref{2}(d), this was done for both the P and AP configurations of the Co electrodes, which can be chosen with the proper magnetic field history. In this case, $H$ was fixed to 250 Oe [see the dots in Fig. \ref{2}(c)], which is large enough to control $M$ of YIG but not to rotate the magnetization of the Co electrodes, as confirmed by magneto-optic Kerr effect (MOKE) microscopy \cite{37,38}. As intended, Fig. \ref{2}(d) shows a clear modulation of the measured $R_{NL}$ (i.e., a modulation of the spin current) when $M$ of YIG is rotated in plane, clearly demonstrating a direct magnetic gating to a pure spin current. The reflection symmetry between the P and AP modulations again rules out the possibility of a relative tilting between the magnetization of Co electrodes \cite{39}. In addition, the measurements were repeated in a control sample, fabricated on a SiO2 substrate, in order to exclude any other possible artifacts \cite{37}. 

\begin{figure}[t]
\centering
\includegraphics[scale=1]{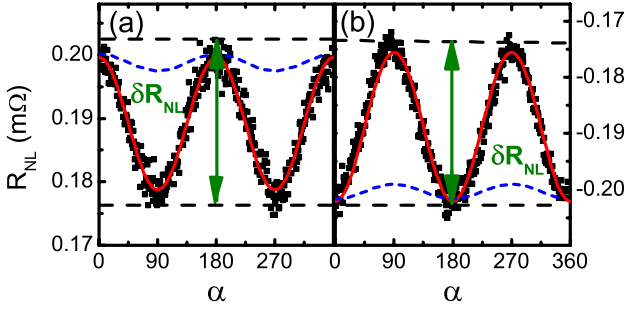}
\caption{\small{(Color online) Nonlocal resistance (black solid squares) as a function of the angle $\alpha$ between the FM electrodes and the applied magnetic field $H$, measured for the parallel (a) and antiparallel (b) configuration, at 150 K and $H=250$ Oe for a LSV with a separation distance of $L=1.6 \text{ }\mu \text{m}$. The red solid line corresponds to the fit of the data to Eq. \ref{eq2}. The blue dashed line corresponds to Eq. \ref{eq2} in the absence of the spin-mixing conductance of the FMI/NM interface. The nonlocal modulation $\delta R_{NL}$ is tagged.}}\label{3}
\end{figure}

The total change in $R_{NL}$, caused by the spin absorption at the Cu/YIG interface, is defined as the nonlocal modulation $\delta R_{NL}=R_{NL}(\alpha=0^{\circ})-R_{NL}(\alpha=90^{\circ})$ (tagged in Fig. \ref{3}). This figure contains the same data from Fig. \ref{2}(d), although, for the sake of clarity, P and AP configurations are plotted separately. In this case, for an $L$ of $1.6 \text{ } \mu \text{m}$, $\delta R_{NL}$ has a magnitude of $\sim 0.025 \text{ m}\Omega$. We can define the factor $\beta=\delta R_{NL}/R_{NL}(\alpha=0^{\circ}$) as an analog of a magnetoresistance, which gives a measure of the efficiency of the magnetic gating. Here, $\beta=8.33\%$ is obtained for the LSV with $L=1.6 \text{ } \mu \text{m}$, whereas $\beta =2.96$\% for $L=570$ nm, showing that longer channels provide more efficient modulations.

In order to quantify the observed modulation of $R_{NL}$, we solve the spin-diffusion equation \cite{20,21,24} in the NM channel,

\begin{equation}
\label{eq1}
\nabla^2\vec \mu_s=\frac{\vec\mu_s}{\lambda^2}+\frac{1}{\lambda_m^2}\vec\mu_s\times{\hat n}\;,
\end{equation}

\noindent where $\vec{\mu}_s$ is the spin accumulation at the NM metal and the vector refers to the spin-polarization direction. $\lambda$ is the spin-diffusion length of the NM and $\lambda_m=\sqrt{\frac{D\hbar}{2\mu_B \mid B\mid}}$ is the magnetic length determined by the amplitude of the magnetic field $B\hat{n}$ ($\hat{n}$ is the unit vector giving its direction). The last term in Eq. \ref{eq1} describes the well-known spin precession due to the applied field \cite{40,41}. $B$ is proportional to $H$ and, for Cu, we can approximate $B\sim\mu_0H$. $D$ is the electronic diffusion constant of the NM, and $\mu_B$ is the Bohr magneton. Assuming $t\ll \lambda$, we can integrate Eq. \ref{eq1} in the $z$ direction and use the Brataas-Nazarov-Bauer boundary condition at the NM/FMI interface \cite{5}. From the solution one can obtain an expression for the nonlocal resistance at the FM detector that reads \cite{37,42,43}

\begin{equation}
\label{eq2}
R_{NL}=\frac{P_I^2R_N}{2}\left[\cos^2\alpha e^{-L/\lambda}+\sin^2\alpha{\rm Re}\left(\frac{\lambda_1}{\lambda}e^{-L/\lambda_1}\right)\right]\;,
\end{equation}

\noindent which is only valid in the high interface resistance limit, i.e., if $R_I\gg R_N$. $P_I$ is the spin polarization of the FM/NM interface at both the FM injector and detector, $R_N=\rho\lambda/wt$ is the spin resistance of the NM, and $\rho$ is its electrical resistivity. The length $\lambda_1$ is defined as 

\begin{equation}
\label{eq3}
\lambda_1=\frac{\lambda}{\sqrt{1+\frac{2\rho G_r\lambda^2}{t}+i\left(\frac{\lambda}{\lambda_m}\right)^2}}\;,
\end{equation}

\noindent where $G_r$ is the real part of the spin-mixing conductance per unit area \cite{5} of the FMI/NM interface. We have disregarded the imaginary part of the spin-mixing conductance in accordance with Refs. \cite{14,32}. Notice that for $\alpha=0^{\circ}$, the $R_{NL}$ for the case without FMI \cite{24,28,29} is recovered: $R_{NL}=\frac{P_I^2 R_N}{2} e^{-L/\lambda}$. At $\alpha=90^{\circ}$ we obtain a similar expression for $R_{NL}$ as in the $\alpha=0^{\circ}$ case, but with a reduced spin-diffusion length ${\rm Re}(\lambda_1)$: $R_{NL}=\frac{P_I^2 R_N}{2} {\rm Re}\left(\frac{\lambda_1}{\lambda} e^{-L/\lambda_1}\right)$. Equation \ref{eq3} shows that two quantities renormalize the spin-diffusion length: the spin-mixing conductance by means of the real term $2\rho G_r\lambda^2/t$, and the imaginary Hanle term $i(\lambda/\lambda_m)^2$ originating from the applied field. While the former leads to a reduction of $\lambda$ due to the torque exerted by the NM/FMI interface to the spins \cite{30,32}, the latter causes, in addition, the precession of the spins when $s$ and $H$ are noncollinear \cite{40}.

\begin{figure}[t]
\centering
\includegraphics[scale=1]{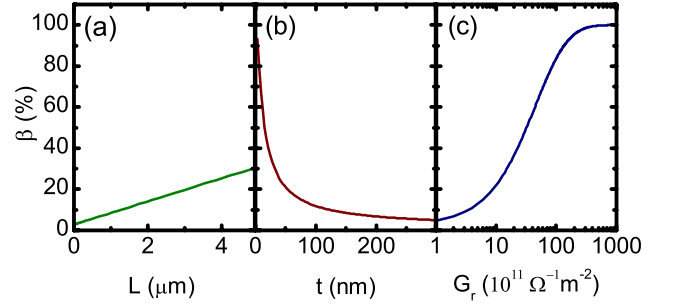}
\caption{\small{(Color online) Representation (solid lines) of the $\beta$ factor, based on Eq. \ref{eq2} for an applied magnetic field $H=250$ Oe, as a function of (a) the distance ($L$) between FM electrodes, (b) the thickness ($t$) of the NM channel, and (c) the spin-mixing conductance per unit area ($G_r$) of the NM/FMI interface. The parameters used for the simulation are:  (a) $\lambda=522$ nm, $\rho=2.1\; \mu\Omega \text{cm}$, $G_r=5\times10^{11}\; \Omega^{-1}\text{m}^{-2}$, and $t=100$ nm. (b) $\lambda=522$ nm, $\rho=2.1\; \mu\Omega \text{cm}$, $G_r=5\times10^{11}\; \Omega^{-1}\text{m}^{-2}$, and $L=1.6\; \mu\text{m}$. (c) $\lambda=522$ nm, $\rho=2.1\; \mu\Omega \text{cm}$, $L=1.6\; \mu\text{m}$, and $t=100$ nm.}}\label{4}
\end{figure}

At a first glance, one might think that the Hanle term could be enough to explain the observed modulation of $R_{NL}$ as a function of $\alpha$. However, as shown in Fig. \ref{3}, a field of 250 Oe in the absence of $G_r$ leads to a modulation of $R_{NL}$ (blue dashed line) which is one order of magnitude smaller than the measured one. This is experimentally confirmed in the control sample performed on top of SiO$_2$ \cite{37}. Increasing $H$ would eventually lead to a Hanle effect of the same order as the $G_r$ effect. Nevertheless, our experiment is limited to low magnetic fields ($H<400$ Oe), to avoid the magnetization of the Co electrodes being affected by the direction of $H$, and thus the Hanle term will not be dominant.

Considering both the $G_r$ and Hanle terms, Eq. \ref{eq2} accurately fits the measured $R_{NL}$ (Fig. \ref{3}), reproducing the observed modulation of the spin current. Note also that Eq. \ref{eq2} reproduces the reflection symmetry between the P and AP configurations, because the product $P_I^2$ has opposite sign for each configuration. The fact that the modulation is observed in a pure spin current in a metal such as Cu excludes any proximity effect as the origin of the modulation \cite{33,34}, confirming the validity of the $G_r$ concept. 

There are two fitting parameters: $P_I$ and $G_r$, whereas $w$, $t$ and $L$ are known geometrical parameters, and $\rho$ (2.1 $\mu\Omega$cm) and $\lambda$ (522 nm) are obtained from resistance measurements and $R_{NL}$ measurements as a function of $L$ \cite{37}. 

From the fitting for the LSV with $L=1.6 \text{ }\mu \text{m}$ (Fig. \ref{3}), we obtained $P_I=0.128 \pm 0.001$ and $G_r=(4.28 \pm 0.06)\times10^{11}\; \Omega^{-1}\text{m}^{-2}$ for the P state [Fig. \ref{3}(a)], and $P_I=0.129 \pm 0.001$ and $G_r=(5.63 \pm 0.07)\times10^{11}\; \Omega^{-1}\text{m}^{-2}$ for the AP state [Fig. (b)], which are almost identical for both magnetic configurations. Therefore, the value of $G_r$ obtained for this particular $L$ is $(4.96 \pm 0.68)\times10^{11}\; \Omega^{-1}\text{m}^{-2}$. The same fitting was performed for the LSV with $L=570$ nm, where it was also possible to measure $R_{NL}$ as a function of $\alpha$, obtaining $P_I=0.123 \pm 0.001$ and $G_r=(2.82 \pm 0.66)\times10^{11}\; \Omega^{-1}\text{m}^{-2}$. Since $G_r$ is extracted separately for each device, this transfers the unavoidable device-to-device dispersion (spin transport is very sensitive to any minor defect) into the value of $G_r$. The difference, which is less than a factor of 2, can thus be considered to be small, taking into account that, in order to observe a relevant variation in $\beta$, a much larger change in $G_r$ is needed [Fig. \ref{4}(c)]. Whereas $P_I$ is within the range reported in similar systems \cite{22,28,29}, $G_r$ is substantially smaller than the values obtained for Pt/YIG (ranging from $1.2\times10^{12}$ to $6.2\times10^{14}\; \Omega^{-1}\text{m}^{-2}$) \cite{6-9,14-16}, Ta/YIG ($4.3\times10^{13}\; \Omega^{-1}\text{m}^{-2}$) \cite{16}, and Au/YIG (between $3.5\times10^{13}$ and $1.9\times10^{14}\; \Omega^{-1}\text{m}^{-2}$) \cite{10,11} either by SMR or spin pumping experiments. 

There is a possibility of underestimating $G_r$ if the assumption for Eq. \ref{eq2}, $R_I \gg R_N$, is not fulfilled. For $\beta\sim8$\%, $G_r$ would increase by a factor of  $\sim2$, to $\sim8 \times10^{11}\; \Omega^{-1}\text{m}^{-2}$, by considering transparent interfaces \cite{37}, which is still low compared to other NM/YIG interfaces. Another possible reason for the low $G_r$ value could be the Ar-ion milling performed before the Cu deposition \cite{12} or the YIG surface quality. We rule this out by performing a control experiment in Pt/YIG where we obtain $G_r=3.34\times10^{13}\; \Omega^{-1}\text{m}^{-2}$ from SMR measurements \cite{37,44,45}. Particularities of the grain structure and the growth condition of the evaporated Cu on YIG could also lead to an effective reduction of $G_r$ at the interface. Alternatively, the spin-orbit interaction effects that might exist for Pt/YIG, Au/YIG or Ta/YIG \cite{35} could lead to an overestimation of the obtained $G_r$ in these systems. Such effects are unlikely in Cu/YIG. It is worth noting that the $G_r$ of a NM/YIG interface, for a NM with a negligible spin-orbit coupling, was not experimentally measured before due to the need of the inverse Spin Hall effect (and thus a high spin-orbit coupling metal) in the experiments made so far \cite{6,7,8,9,10,11,12,13,14,15,16}.

Finally, a representation of $\beta$, based on Eq. \ref{eq2}, is plotted in Fig. \ref{4} as a function of different parameters ($L$, $t$ and $G_r$) which can be controlled in order to improve the efficiency of the magnetic gating. The values of the different parameters used for the representation are listed in the caption and correspond to realistic values taken from our devices. $\beta$ increases linearly with the length ($L$) between the FM electrodes, reaching $\sim30$\% for $L=5\; \mu\text{m}$ [Fig. \ref{4}(a)]. When the spin current flows over a longer distance, the spin scattering and absorption caused by the NM/FMI interface will be enhanced (i.e., $\beta$ will be larger). This is in agreement with our experimental results discussed above. However, there is an experimental limit, since the nonlocal signal decays exponentially and will be negligible when $L\gg \lambda$ \cite{23,26}. By decreasing the thickness ($t$) of the Cu channel, $\beta$ increases asymptotically when $t$ approaches 0 [Fig. \ref{4}(b)]. In this case, by decreasing $t$, the relative contribution of the NM/FMI interface to the spin-flip scattering processes increases, enhancing $\beta$. For instance, when $t\sim20$ nm, $\beta$ already increases to $\sim50$\%. However, the decrease of $\lambda$ with $t$ \cite{26}, which has not been taken into account for the representation, will lower $\beta$. The most effective way of improving $\beta$ seems to be increasing $G_r$ [Fig. \ref{4}(c)]. By increasing it by two orders of magnitude, i.e., for a $G_r$ of the order that Pt/YIG systems have, $\beta$ reaches almost up to a 100\%, which would lead to a perfect magnetic gating of the pure spin currents. This seems feasible by improving the interface between Cu and YIG or by using another NM material with a high spin-mixing interface conductance with YIG. 

To conclude, we present an approach to control and manipulate spins in a solid state device, by means of a magnetic gating of pure spin currents in Co/Cu LSV devices on top of YIG. A modulation of the pure spin current is observed as a function of the relative orientation between the magnetization of the FMI and the polarization of the spin current. Such modulation is explained by solving the spin-diffusion equation and considering the spin-mixing conductance at the NM/FMI interface. The accuracy between the measured data and the expected modulation provides an effective way of studying the NM/FMI interface. From our results, a spin-mixing conductance of $G_r\sim4\times10^{11}\; \Omega^{-1}\text{m}^{-2}$ is obtained for the Cu/YIG interface. An increase of this value will enhance the efficiency of the magnetic gating. This can be achieved by carefully tuning the fabrication parameters.  Our experiment paves the way for different manners of spin manipulation, bringing closer pure spin currents and logic circuits.

\vspace{1cm}

\noindent \textbf{ACKNOWLEDGEMENTS}

\vspace{0.5cm}

We thank Professor Joaqu\'in Fern\'andez-Rossier for fruitful discussions. This work was supported by the European Commission under the Marie Curie Actions (256470-ITAMOSCINOM), NMP Project (263104-HINTS) and the European Research Council (257654-SPINTROS), by the Spanish MINECO under Projects No. MAT2012-37638, No. MAT2012-36844, and No. FIS2011- 28851-C02-02, and by the Basque Government under Project No. PI2012-47 and UPV/EHU Project No. IT-756-13. E.V. and M.I. thank the Basque Government for support through a Ph.D. fellowship (Grants No. BFI-2010-163 and No. BFI-2011-106). F.S.B. thanks Professor Martin Holthaus and his group for their kind hospitality at the Physics Institute of the Oldenburg University.


\clearpage

\begin{widetext}
\setcounter{figure}{0}
\renewcommand{\thefigure}{S\arabic{figure}}
\setcounter{equation}{0}
\renewcommand{\theequation}{S\arabic{equation}}
\newcounter{multibblnoresetbibitemcount}

\linespread{1.5}
\begin{center}
\Large{\textbf{Modulation of pure spin currents with a ferromagnetic insulator}}
\\
\Large{\textbf{SUPPLEMENTAL MATERIAL}}
\\*
\vspace{0.6cm}
\large{Estitxu Villamor, Miren Isasa, Sa\"ul V\'elez, Amilcar Bedoya-Pinto,}
\\*
\large{Paolo Vavassori, Luis E. Hueso, F. Sebasti\'an Bergeret, and F\`{e}lix Casanova}
\end{center}

\linespread{1.5}


\vspace{0.8cm}

\noindent{\textbf{S1. Experimental details}}

\vspace{0.5cm}

Lateral spin valves (LSVs) were fabricated by a two-step electron-beam lithography, ultra-high-vacuum evaporation and lift-off process (see Fig. \ref{2}(b) from the main text for a SEM image of the device). Since yttrium iron garnet (YIG) on gadolinium gallium garnet (GGG) is an insulating substrate, a thin gold (Au) layer of ~2.5 nm was sputtered on top of the PMMA resist before each lithography step. This prevents the charging of the substrate during the e-beam exposure, which otherwise would distort the pattern. After the e-beam exposure and before developing the patterned resist, the Au layer was removed with Au etchant. In the first lithography step, FM electrodes were patterned and cobalt (Co) was electron-beam evaporated with a base pressure $\leq1\times10^{-9}$ mbar. In the second lithography step, the NM channel was patterned and copper (Cu) was thermally evaporated with a base pressure $\leq1\times10^{-9}$ mbar. Ar-ion milling was performed prior to the Cu deposition in order to remove resist left-overs; the parameters used for the Ar-ion milling are an Ar flow of 15 standard cubic centimeters per minute, an acceleration voltage of 50 V, a beam current of 50 mA, and a beam voltage of 300 V for 30 s, as described in Ref. \cite{s1}. To overcome the low spin injection of Co when using transparent interfaces \cite{s1,s2}, an oxide layer was created at the Co/Cu interface by letting the Co oxidize after the milling and before the Cu deposition. The presence of an interface resistance is known to enhance the spin injection efficiency \cite{s3}. An interface resistance $R_I\geq 5 \text{ }\Omega$ is estimated in this case. Both Co electrodes have a thickness of 35 nm and different widths (115 nm and 175 nm) to obtain different switching fields by means of shape anisotropy. Three LSVs were fabricated, bridged by the same Cu wire (of width $w \sim200 \text{ nm}$ and thickness $t \sim100 \text{ nm}$) with edge-to-edge distances between the Co electrodes of $L$ = 250 nm, 570 nm and 1600 nm. 

Such configuration allows the measurement of the four-point resistance $R$ of the wire as a function of $L$ (where the electrodes belong to the same LSV or to two contiguous LSVs). By performing a linear fitting of $R$ as a function of $L$ (see lower inset in Fig. \ref{S1}) and knowing the dimensions of the Cu wire, it is straightforward to obtain its electrical resistivity, which has the value of $\rho = 2.1\text{ } \mu\Omega \text{cm}$ at a temperature of 150 K.

The non-local resistance $R_{NL}=V_{NL}/I$, i.e. the non-local voltage $V$ measured at the detector normalized to the value of the injected current $I$, is measured as a function of the applied magnetic field $H$ for the three LSVs.  Note that the LSV with $L$ = 250 nm broke after measuring $R_{NL}$ as a function of $H$ at $\alpha=0^\circ$ and, for this reason, the results of the $R_{NL}$ measurement as a function of $\alpha$ are not included in the main text. $R_{NL}$ changes from positive to negative when the relative magnetization of the FM electrodes changes from parallel (P) to antiparallel (AP) by sweeping $H$ (see upper inset of Fig. \ref{S1}). The difference between positive and negative $R_{NL}$ is defined as the spin signal ($\Delta R_{NL}$), which, for LSVs with a high interface resistance, can be expressed as \cite{s3,s4,s5,s6}:

\begin{eqnarray}
\Delta R_{NL}=P_I^2R_Ne^{-L/\lambda}\; ,\label{eqs1}
\end{eqnarray}

\noindent where $P_I$ is the spin polarization of the Co/Cu interface, $R_N=\rho\lambda/wt$ is the spin resistance of Cu and $\lambda$ is the spin diffusion length of Cu. Figure \ref{S1} shows the measured $\Delta R_{NL}$ as a function of $L$ (black squares), which is fitted to Eq. (\ref{eqs1}) (red solid line) in order to obtain $P_I$ and $\lambda$. The obtained values at 150 K are $P_I = 0.18\pm 0.01$ and $\lambda = 522 \pm 25\text{ nm}$. 
Even though $P_I$ is within the range of values that are observed in literature in similar systems \cite{s2,s4,s5,s6,s7}, it is slightly higher than the $P_I$ values obtained in the main text for the fitting of Eq. \ref{eq2}. This is due to the dispersion of the interface quality between different LSVs. The device with $L = 250 \text{ nm}$ has a higher $P_I$, which enhances the averaged $P_I$ obtained from the fitting of Eq. (\ref{eqs1}).
The value obtained for $\lambda$ is similar but slightly lower than our previous values obtained in Py/Cu LSVs on top of $\text{Si/SiO}_2$ measured at 150 K ($\lambda = 680\pm15 \text{ nm}$) \cite{s1}. This could be due to the different growth of Cu on top of YIG as compared to $\text{SiO}_2$.

\begin{figure}[h]
\centering
\includegraphics[scale=1]{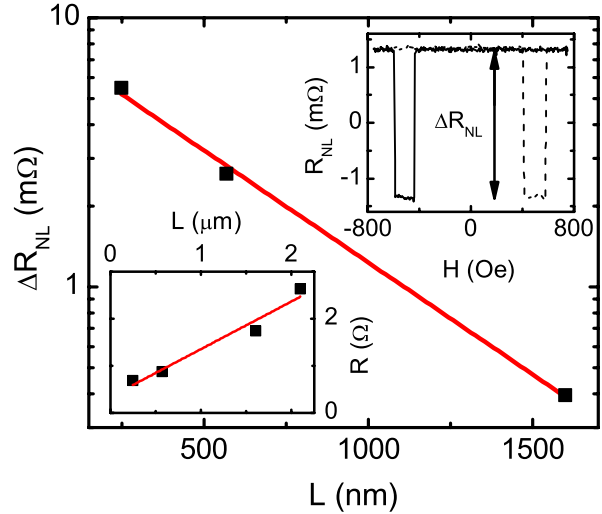}
\caption{\small{Spin signal as a function of the separation distance between Co electrodes ($L$) for three Co/Cu LSVs, measured at 150 K with the applied field ($H$) parallel to the electrodes (black squares). Red line is a fit to Eq. (\ref{eqs1}). Upper inset: Non-local resistance measured as a function of $H$ for the LSV with $L = 570 \text{ nm}$. A constant background of 0.38 m$\Omega$ is subtracted from the data. Solid (dashed) line indicates the decreasing (increasing) sweep of $H$. Spin signal $\Delta R_{NL}$ is tagged in the plot. Lower inset: Four-point resistance as a function of $L$ (black squares). Red line is a linear fit.}}\label{S1}
\end{figure}

\newpage
\noindent{\textbf{S2. Control experiment: MOKE measurements to rule out magnetization rotation in Co electrodes}}
\vspace{0.5cm}

A tilting of the magnetization direction in the Co electrodes during the measurement of $R_{NL}$ as a function of the angle $\alpha$ between the Co electrodes ($y$ direction, see Fig. \ref{1} from the main text) and the direction of the applied magnetic field $H$, could in principle be invoked to explain the observed modulation of $R_{NL}$ with $\alpha$ because $R_{NL}\propto \cos\theta$, where $\theta(\alpha)$ is the relative angle between the magnetizations of both electrodes. This tilting could be caused by the torque exerted directly by the applied magnetic field or/and by a coupling between the Co electrodes and the YIG substrate. A modulation of $\sim$ 8\%, such as the observed one, could correspond to $\theta\sim23^{\circ}$. Even if the reflection symmetry between the $R_{NL}$ modulation observed in the parallel and anti-parallel magnetization states of the Co electrodes is sufficient to rule out such explanation (as stated in the main text), to further exclude a magnetization rotation of the electrodes, MOKE microscopy measurements were performed at room temperature directly on the same sample used for the magnetic gating experiment. The capability of our MOKE microscope  to measure the field induced magnetization reorientation of ultra-small ferromagnetic nanostructures was demonstrated earlier \cite{s8}. The MOKE measurements were performed on top of the widest electrode, which is the one whose magnetization can rotate more easily due to shape anisotropy.

Figure \ref{S2} shows hysteresis loops of the Co electrode (red circles) and of the YIG (black squares), \emph{i.e.} the projection of the magnetization in the $y$ direction, $M_y$, is measured as a function of the magnetic field applied in the $y$ direction, $H_y$, and normalized to the saturation magnetization, $M_s$. In both cases,the MOKE signal was acquired from a subset of the pixels of the CCD detector that corresponds to an area on the sample surface equal to $100\times800$ nm$^2$ \cite{s8}. The coercive field of the Co electrode is 500 Oe, in agreement with the $R_{NL}$ measurements as a function of $H$ shown in Fig. \ref{1}(c) from the main text. For the YIG substrate, magnetic saturation around 100 Oe is observed, in agreement with the VSM measurements shown in Fig \ref{1}(a) from the main text.

\begin{figure}[h]
\centering
\includegraphics[scale=0.3]{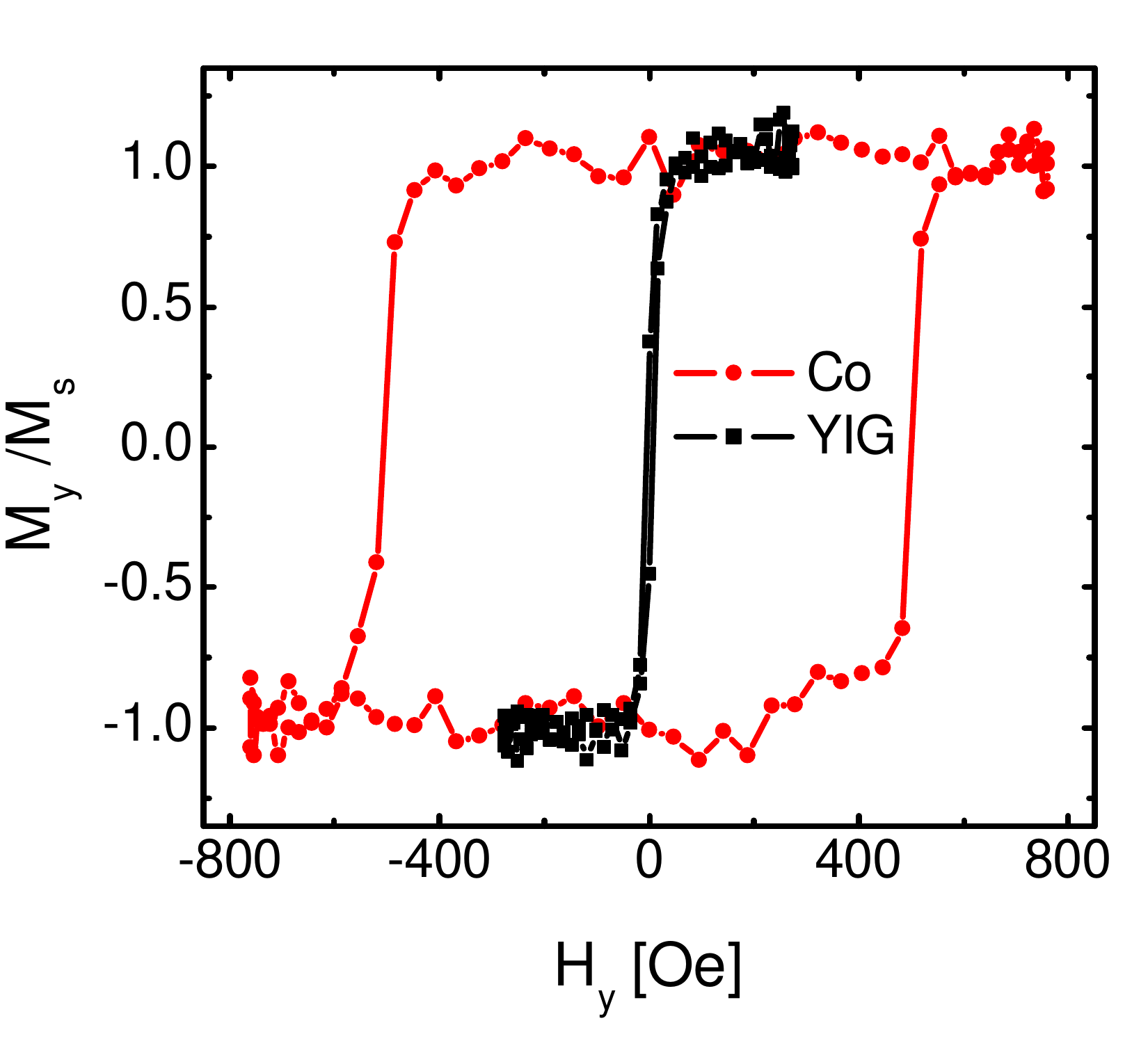}
\caption{\small{Projection of the magnetization in the $y$ direction, $M_y$, of the YIG (black squares and line) and of  the Co electrodes (red circles and line) normalized to the saturation magnetizations, $M_s$, measured as a function of the magnetic field applied in the $y$ direction, $H_y$. Measurements are performed at 300 K.}}\label{S2}
\end{figure}

To check for a possible rotation of the magnetization of the Co electrode, its $M_y/M_s$ was measured while the direction of the magnetic field $H$, which had a fixed intensity of 250 Oe, was rotated by $\alpha$ varied from 0 to $360^{\circ}$. Figure \ref{S3} shows $M_y/M_s$ of the Co electrode and the YIG substrate. Whereas the magnetization of YIG coherently rotates with the direction of $H$ ($M_y/M_s\propto\cos\alpha$), given that $H$ is largely exceeding the saturation field of YIG, the magnetization of the Co electrode is constant for every $\alpha$. Based on the signal-to-noise ratio of our measurements, the smallest detectable change in $M_y/M_s$ corresponds to a rotation of $M_s$ of less than 5$^{\circ}$. Therefore, from our measurements, we can directly conclude that the rotation of the Co magnetization, if any, is less than $5^{\circ}$, which could only explain a variation of less than 0.4\% in $R_{NL}$, well below the experimentally observed $\sim$8\% .

\begin{figure}[h]
\centering
\includegraphics[scale=0.3]{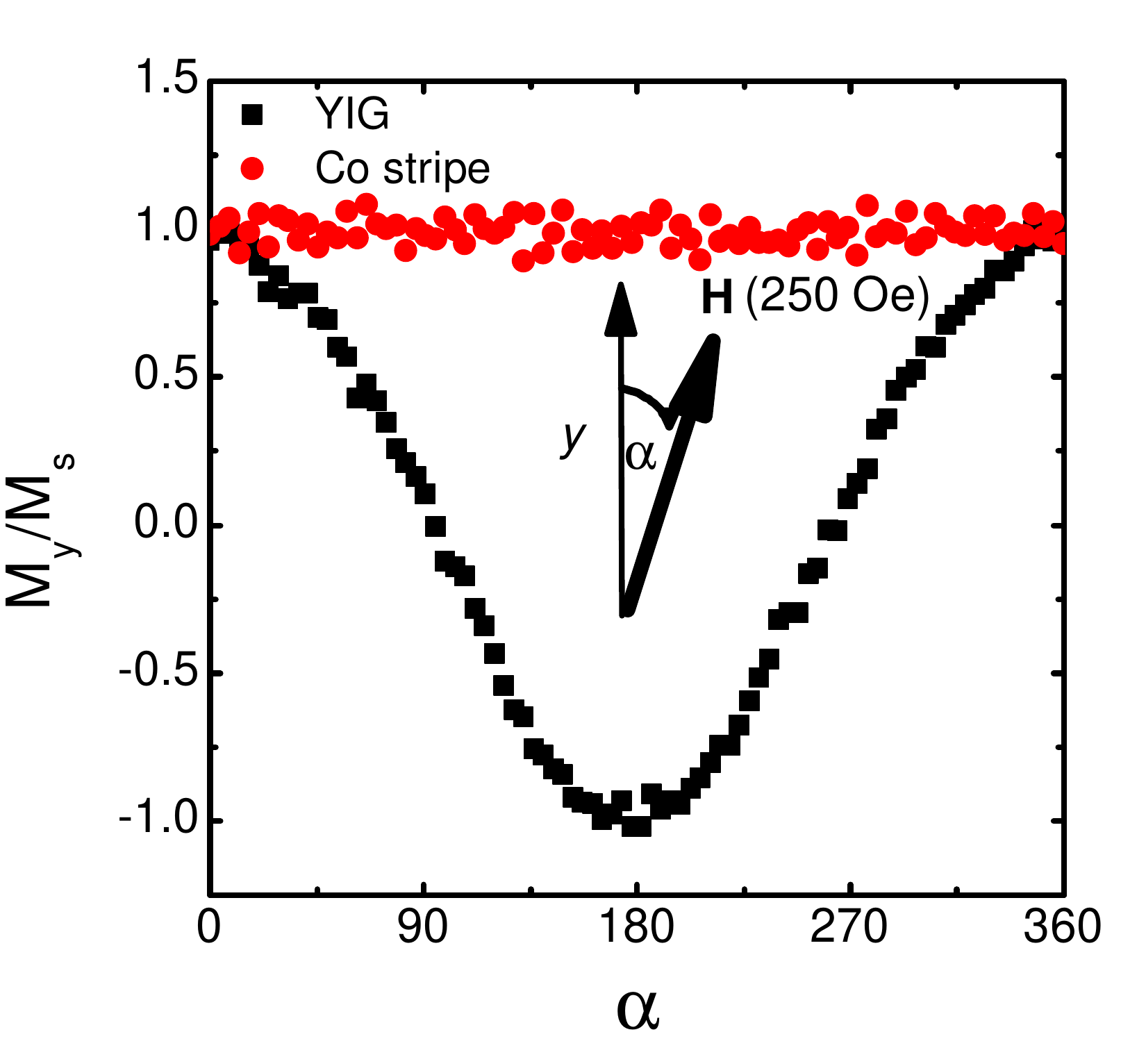}
\caption{\small{Projection of the magnetization in the $y$ direction, $M_y$, of the YIG (black squares) and of  the Co electrodes (red circles) normalized to the saturation magnetizations, $M_s$, measured as a function of the angle $\alpha$ between the direction of the Co electrodes ($y$) and the applied magnetic field, $H = 250$ Oe. Measurements are performed at 300 K.}}\label{S3}
\end{figure}

\clearpage

\noindent{\textbf{S3. Control experiment: Non-local resistance measurements on top of a silicon oxide substrate}}
\vspace{0.5cm}

Even though a possible rotation of the FM electrodes, which could explain a variation in the non-local resistance as the one we observe, was excluded with the previous control experiment (section S2), an additional control experiment was performed in order to rule out any other possible artifact. With this purpose, the main experiment was repeated in a LSV fabricated on top of SiO$_2$ instead of YIG. Fig \ref{S4}(b) shows $R_{NL}$ measured at 150 K, applying a magnetic field of 250 Oe, as a function of $\alpha$ for both the parallel (P) and antiparallel (AP) magnetizations of the FM electrodes in a LSV with $L=750$ nm. Apparently, no periodic modulation of $R_{NL}$ is observed. However, by taking a closer view (Figs. \ref{S4}(c) and \ref{S4}(d)), one can guess a periodic modulation of the order of the noise, which behaves as the one observed in the main experiment, including a reflection symmetry between the P and AP case. The observed modulation corresponds to a $\sim1$\%, which is the estimated value for the Hanle effect at this $L$, and is certainly smaller than the $\beta=2.96$\% and $\beta=8.33$\% observed in the main experiment for $L=570\ \text{nm}$ and $L=1.6\ \mu$m, respectively. This confirms experimentally that, for these values of $L$ and $H$, the Hanle contribution is much smaller than the modulation due to spin-mixing conductance, and also excludes any other possible artifact.


\begin{figure}[h]
\centering
\includegraphics[scale=1.3]{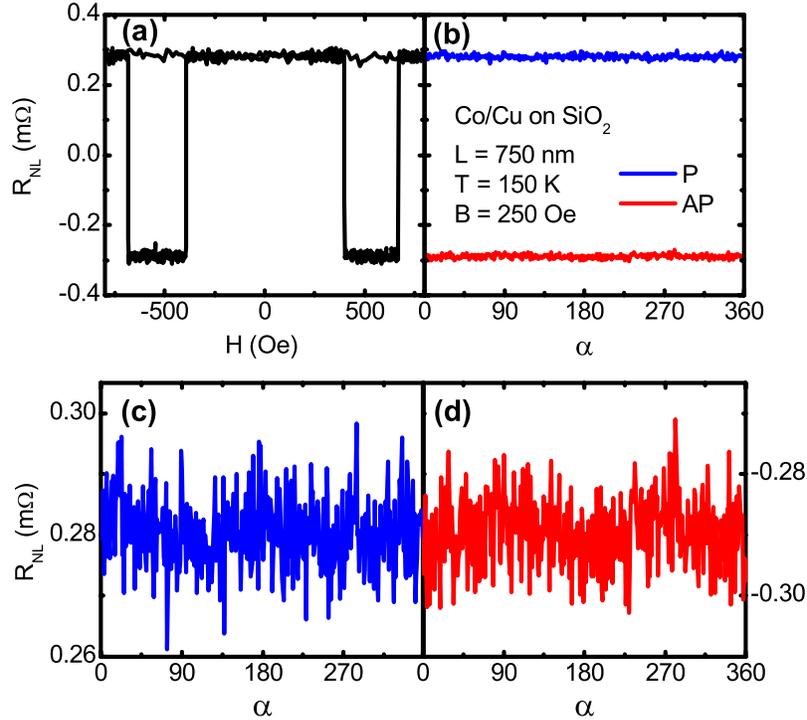}
\caption{\small{(a) Non-local resistance ($R_{NL}$) measured at 150 K as a function of $H$ with $\alpha=0^{\circ}$ for a LSV fabricated on top of SiO$_2$ with $L=750$ nm. (b) $R_{NL}$ as a function of $\alpha$ measured for both the (c) P and (d) AP configuration, at 150 K with $H=250$ Oe for the same LSV.}}\label{S4}
\end{figure}
\newpage
\noindent{\textbf{S4. Control experiment: Spin Hall magnetoresistance measurements in a Pt/YIG sample}}
\vspace{0.5cm}

In order to see if the low $G_r$ value obtained for Cu/YIG interfaces originates from the quality of the YIG substrate or from any effect that might be induced at the YIG substrate for the Ar-ion milling process (see section S1), we fabricated a Pt/YIG control sample and tested it within the Spin Hall magnetoresistance (SMR) framework. The SMR arises from the simultaneous effect of the spin Hall effect and the inverse spin Hall effect in the Pt layer in combination with the interaction of the generated spin current with the magnetization of the YIG surface. Depending on the relative orientation between the spin polarization vector and the direction of the magnetic moment of the YIG surface, the spin current might be absorbed via spin-transfer torque resulting in a modulation of the resistance of the Pt layer, which is fundamentally related to the spin-mixing conductance $G_r$ of the Pt/YIG interface \cite{s9,s10}. 

With this purpose, a 7-nm-thick Pt Hall bar (with a width $W = 100\text{ }\mu\text{m}$ and a length $L = 800\text{ }\mu\text{m}$) was sputtered on top of a YIG substrate grown as the one used for the fabrication of the LSV. Prior to the Pt deposition, the YIG surface was subjected to the same Ar-ion milling process (see section S1). Angular dependent magnetoresistance (ADMR) measurements were performed by rotating a fixed $H$ along the three main rotation planes of the system: XY, YZ and XZ, with the corresponding angles $\alpha$, $\beta$ and $\gamma$.  A large enough $H$ is applied to ensure the magnetization of the YIG substrate follows the direction of the applied magnetic field. The resistance was measured using both longitudinal ($R_L$) and transverse ($R_T$) configurations. Figure \ref{S5}(a) shows a sketch of the resulting device with the definition of the axes and the transverse configuration. As expected from the SMR theory \cite{s9,10}: (\emph{i}) no ADMR is observed in $R_L(\gamma)$, (\emph{ii}) a large modulation is observed in $R_L(\alpha)$ and $R_L(\beta)$, with the same amplitude and a $\text{cos}^2$ dependence, and (\emph{iii}) $R_T(\alpha)$ shows a $\sin\alpha\cdot\cos\alpha$ dependence, with the same amplitude as in $R_L(\alpha)$ but with a $L/W$ factor. As illustrative example, the transverse resistance $R_T(\alpha)$ obtained for $H$ = 1 kOe and 150 K is plotted in Fig. \ref{S5}(b). 

\begin{figure}[h]
\centering
\includegraphics[scale=1]{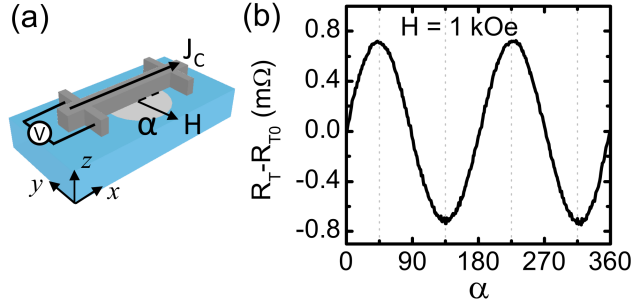}
\caption{\small{(a) Sketch of the Pt Hall bar on YIG. Charge current ($J_C$) and applied external magnetic field ($H$), measurement configuration, axes and the angle ($\alpha$) between $H$ and $J_C$ are indicated. (b) Transverse resistance ($R_T$) measured as a function of $\alpha$. A small spurious baseline resistance $R_{T0}$ was subtracted.}}\label{S5}
\end{figure}

According to the SMR theory, the amplitude of the observed magnetoresistance is related to the microscopic properties of the Pt layer by \cite{s9,s10}: 

\begin{eqnarray}
 \frac{\Delta\rho}{\rho}=\theta_{SH}^2\frac{\frac{2\rho\lambda^2}{t}G_r \tanh^2\frac{t}{2\lambda}}{1+2\rho\lambda G_r \coth\frac{t}{\lambda}}  \; ,\label{eqs2}
\end{eqnarray}

\noindent where $\theta_{SH}$ is the spin Hall angle, $\lambda$ is the spin diffusion length, $\rho$ is the electrical resistivity, $t$ is the thickness of the Pt and $G_r$ is the real part of the spin-mixing conductance per unit area of the Pt/YIG interface. In our case, with a measured longitudinal resistance of $R_L = 281.5 \text{ }\Omega$ at 150 K, one can determine $\rho = 24.7 \text{ }\mu\Omega \text{ cm}$ and the SMR signal $\Delta\rho/\rho=5.48\times10^{-5}$. Knowing the values of $\theta_{SH}$ and $\lambda$ in Pt, one can extract the $G_r$ value of the Pt/YIG interface using Eq. \ref{eqs2}. These values cannot be inferred from our measurements, but can be obtained from literature. Despite there is a big dispersion of the given values for $\theta_{SH}$ and $\lambda$ \cite{s9, s11, s12}, we will use the ones recently reported in Ref. \cite{s12} ($\theta_{SH}$  = 0.056 and $\lambda$ = 3.4 nm), since they are highly consistent within different methods used to determine them. A $G_r = 3.4\times10^{13} \text{ }\Omega^{-1}m^{-2}$ is thus obtained for our Pt/YIG interface, which is in agreement with the previously reported range of values \cite{s9,s13,s14,s15,s16,s17,s18}. We can take this result as a proof of the good quality of the YIG substrates used in the present experiments and as an indication that the Ar-ion milling process in the LSV experiment is not at the origin of the low $G_r$ obtained.

\vspace{0.5cm}
\noindent{\textbf{S5. Theory}}
\vspace{0.5cm}

In order to model the experimental results, we consider the geometry shown in Fig. \ref{1} from the main text. It consists of a normal (NM) layer, Cu in our case, deposited on top of a ferromagnetic insulator (FMI), YIG in this case. 
At $x=0$ there is a ferromagnetic (FM) electrode, Co in our case, that injects a charge current $I$ that flows in $x<0$ direction.  Coming from a FM, this current induces a spin accumulation $\mu_s^a$ ($a$ denotes the spin polarization direction) in the NM layer.  In the absence of spin-orbit coupling, a spin current density in the NM $j_k^a$ ($k$ denotes the flow direction) is then originated by the gradient of the spin accumulation $\mu_s^a$
\begin{equation}
\label{curr_den}
j_k^a=-\frac{1}{2e\rho}\partial_k\mu_s^a\; ,
\end{equation}
where $\rho$ is the electrical  resistivity  of the NM and $e$ the absolute value of the electron charge. In a normal metal, the mean free path $l$ is smaller than other characteristic lengths, and therefore the spin accumulation is determined by the Bloch equation with an added diffusion term \cite{s19, s20}, which, in the steady state, has the simple form:
\begin{equation}
\label{diffusion}
\nabla^2\vec \mu_s=\frac{\vec\mu_s}{\lambda^2}+\frac{1}{\lambda_m^2}\vec\mu_s\times{\hat n}\; .
\end{equation}
Here  $\vec {\mu_s}=(\mu_s^x,\mu_s^y,\mu_s^z)$, $\lambda$ denotes the  spin diffusion length which is related to the diffusion coefficient $D$ and the spin-flip relaxation time $\tau_{sf}$ by $\lambda=\sqrt{D\tau_{sf}}$, and $\lambda_m=\sqrt{D\hbar/2\mu_BB}$ is the magnetic length determined by the amplitude of the applied magnetic field $B{\hat n}$ (${\hat n}$ is a unit vector along the magnetic field direction).
Alternatively, Eq. (\ref{diffusion}) can be derived from the Keldysh Green's function formalism \cite{s21,s22}. 
In the case of an intrinsic spin-orbit coupling, due to an inversion asymmetry, Eqs. (\ref{curr_den}-\ref{diffusion}) have the same form if one substitutes the gradient by a SU(2) covariant derivative \cite{s21}. In the case of extrinsic spin-orbit coupling, due to random impurities, these equations acquire some extra terms \cite{s22}. However, spin-orbit effects are negligible in accordance in our NM, Cu. 

We assume that the system is invariant in $y$ direction and therefore the spin accumula-\\tion only depends on $x$ and $z$: $\mu_s(x,z)$. In order to solve the diffusion equation (\ref{diffusion}) for the spin accumulation one needs proper boundary conditions. At the upper interface of the NM with the vacuum the spin current should vanish:
\begin{equation}
\label{bc_vac}
\partial_z\mu_s|_{z=t}=0\;,
\end{equation}
where $t$ is the thickness of the NM. We are assuming $z=0$ at the NM/FMI interface, and $z=t$ at the NM/vacuum interface. At the interface with the FMI we use the Brataas-Nazarov-Bauer boundary condition \cite{s23}:
\begin{equation}
\label{bc_yig}
\partial_z\vec\mu_s|_{z=0}=-2\rho\left[G_r {\hat m}\times({\hat m}\times\vec \mu_s)+G_i{\hat m}\times\vec \mu_s\right]\;, 
\end{equation}
where ${\hat m}$ is a unit vector along the magnetization of the FMI, and $G_{mix}=G_r+iG_i$ is the the complex spin-mixing interface conductance per unit area \cite{s23}. In the experiment the thickness $t$ of the NM layer is smaller than the characteristic scale of variation of $\mu_s$ ($\approx \lambda$) and therefore we can integrate Eq. (\ref{diffusion}) over $z$ assuming that $\mu_s$ does not depend on $z$. By performing this integration and  using Eqs. (\ref{bc_vac}-\ref{bc_yig}) we obtain the following (1D) equation for $\vec \mu_s$:
 \begin{equation}
 \label{1Deq}
\partial^2_{xx}\vec\mu_s=\frac{\vec\mu_s}{\lambda^2}+\left( \frac{1}{\lambda_m^2}+\frac{1}{\lambda_i^2}\right)\vec\mu_s\times{\hat m}-\frac{1}{\lambda_r^2}{\hat m}\times({\hat m}\times\vec \mu_s)\; ,
 \end{equation}
where we have considered an in-plane magnetization of the FMI, ${\hat m}=(\sin\alpha,\cos\alpha,0)$, and defined $\lambda_r^{-2}=2\rho G_r/t$ and $\lambda_i^{-2}=2\rho G_i/t$. The latter term acts as an effective magnetic field parallel to the magnetization of the FMI which is assumed to be parallel to the applied magnetic field.

Equation (\ref{1Deq}) describes the spatial dependence of the spin accumulation in a thin FM layer in contact with a FMI. It consists of three coupled linear second order differential equations.  In order to solve it we have to write the boundary conditions corresponding to the experimental situation: at $x=0$ an electric current $I$ is injected. This induces at $x=0$ a spin current equal to $P_II$, where  $P_I$ is the spin polarization of the FM/Cu interface. At a distance $L$ from the injection point there is a detector. The spin accumulation and its derivative are continuous in the NM layer. The boundary conditions for $\vec \mu_s(x)$ at the injector and detector are obtained from the spin current continuity and read:
\begin{eqnarray}
\label{bc_inj}
P_II{\hat y} &=&- \frac{\lambda}{eR_N}\partial_x\vec \mu_s|_{x=0^-}- \frac{\lambda}{eR_N}\partial_x\vec\mu_s |_{x=0^+}\\
\label{bc_det}
0&=&- \frac{\lambda}{eR_N}\partial_x\vec \mu_s|_{x=L^-}- \frac{\lambda}{eR_N}\partial_x\vec\mu_s |_{x=L^+}\; ,
\end{eqnarray}
where the spin current at both sides of the FM injector (detector) is considered.  $R_N=\rho\lambda/wt$ is defined as the spin resistance, where $w$ is the width of the NM channel. The FM injector is polarized in $y$ direction (whose unit vector is ${\hat y}$) due to shape anisotropy, and, thus, the injected spin current as well. In order to obtain the boundary conditions Eqs. (\ref{bc_inj}-\ref{bc_det}), a high interface resistance ($R_I$) was considered at the interfaces between the NM and the FM injector ($x = 0$) and between the NM and the FM detector ($x=L$) \cite{s3}, i.e. $R_I\gg R_N$. If  $R_I$ is of the order of $R_N$, a spin current that might flow back into the FM electrodes \cite{s20,s24} has to be taken into account. 

In the case considered above, it is rather straightforward to solve Eq. (\ref{1Deq}) with the conditions Eqs. (\ref{bc_inj}-\ref{bc_det}), in order to obtain the spin accumulation in all three spin polarization directions:
\begin{eqnarray}
\mu_s^x(x)&=&P_IIeR_N\cos\alpha\sin\alpha\left[e^{-x/\lambda}+{\rm Re}\left(\frac{\lambda_1}{\lambda}e^{-x/\lambda_1}\right)\right]\label{sol_mux}\, ,
\\\mu_s^y(x)&=&P_IIeR_N\left[\cos^2\alpha e^{-x/\lambda}+\sin^2\alpha{\rm Re}\left(\frac{\lambda_1}{\lambda}e^{-x/\lambda_1}\right)\right]\label{sol_muy}\, ,
\\\mu_s^z(x)&=&-P_IIeR_N\sin\alpha{\rm Im}\left(\frac{\lambda_1}{\lambda}e^{-x/\lambda_1}\right)\label{sol_muz}\, ,
\end{eqnarray}
where the characteristic length in the second exponential is defined as
\begin{equation}
\label{lambda1}
\lambda_1=\frac{\lambda}{\sqrt{1+\gamma}}
\end{equation}
with $\gamma=\gamma_r+i\gamma_i$, $\gamma_r=\lambda^2/\lambda_r^2$ and $\gamma_i=\lambda^2(1/\lambda_m^2+1/\lambda_i^2)$.  

It is interesting to note that, even if the injected spin current is polarized in the $y$ direction, a spin accumulation is created with the spins polarized in the $x$ direction, due to both the torque exerted by $M$ at the NM/FMI interface and the spin precession caused by the magnetic field perpendicular to the spin polarization, and in the $z$ direction only due to the spin precession caused by the magnetic field perpendicular to the spin polarization. 

Since the magnetization of the injector and detector are in $y$ direction, only  $\mu_s^y$ can be detected at $x=L$. From Eq. (\ref{sol_muy}) we can determine the non-local resistance ($R_{NL}$)  defined in terms of the non-local voltage $V_{NL}$ measured at the detector \cite{s3,s24}:
\begin{equation}
\label{def_rnl}
R_{NL}=\frac{V_{NL}}{I}=\frac{P_I\mu_s^y(L)}{2eI}\; ,
\end{equation}
where we assume that the polarization at the detector contact is the same as at the injector. By inserting Eq. (\ref{sol_muy}) into this last expression we finally obtain
\begin{equation}
\label{sol_rnl}
R_{NL}=\frac{P_I^2R_N}{2}\left[\cos^2\alpha e^{-L/\lambda}+\sin^2\alpha{\rm Re}\left(\frac{\lambda_1}{\lambda}e^{-L/\lambda_1}\right)\right]\; .
\end{equation}
It follows that, in the absence of the spin-mixing conductance and if the magnetic field is in the $x$ direction ({\it i.e.} $\alpha=90^{\circ}$), the expression is identical to the one obtained in Ref. \cite{s20} in which  the Hanle effect was studied.  

Eq. (\ref{sol_rnl}) is a general expression that describes the non-local resistance in the NM/FMI structure and takes into account both the effect of the external applied field and the spin-mixing conductance describing the magnetic interactions at the interface. We have fitted our measurements of the  $R_{NL}(\alpha)$ dependence with Eq. (\ref{sol_rnl}) by neglecting the imaginary part of the spin-mixing conductance which, according to first-principle calculations \cite{s25} and our discussion below, seems to be a good approximation. As we can see in Fig. \ref{3} from the main text, the effect of the applied field on the $R_{NL}(\alpha)$ modulation is small in comparison to the one induced by $G_r$. This demonstrates that the modulation observed can only be explained by the effect of the spin-mixing conductance.  According to our estimations, $G_r\approx4\times10^{11}\text{ }\Omega^{-1}m^{-2}$. This value is in principle smaller than the value reported in previous works \cite{s9,s13,s14,s15,s16,s17,s18,s26,s27}. This discrepancy is discussed in the main text.

\begin{figure}[h]
\centering
\includegraphics[scale=1]{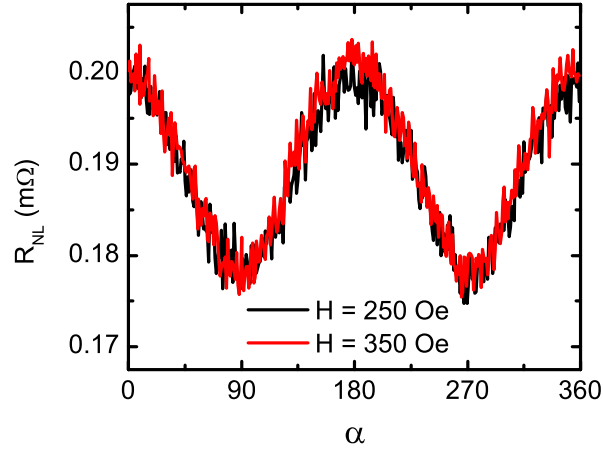}
\caption{\small{Non-local resistance measured as a function of the angle $\alpha$ between the spin polarization and the applied magnetic field $H$. Two measurements have been done for $H=250\text{ Oe}$ and $H=350\text{ Oe}$, with identical results.}}\label{S6}
\end{figure}

The experimental results shown in the main text have been obtained for an applied magnetic field of 250 Oe. Measurements have been performed also at 350 Oe with almost identical results (see Fig. \ref{S6}).  From theory, the  difference in $R_{NL}$ for these two fields is of the order of the measurement noise and, thus, not detectable in principle. For such values, $(\lambda/\lambda_m)^2\ll 1$.  But  also $(\lambda/\lambda_r)^2$ (and presumably also  $(\lambda/\lambda_i)^2$)  are very small according to our estimation of $G_r$. For a $G_r\approx4\times10^{11}\text{ }\Omega^{-1}m^{-2}$, and with the parameters of the used LSVs, $\left(\lambda/\lambda_r\right)^2=0.037$ is obtained. Therefore, one can go analytically one step further by treating the parameter $\gamma$ in Eq. (\ref{lambda1}) perturbatively. 

Instead of focusing in $R_{NL}$ let us analyze the amplitude of the effect when varying  $\alpha$ (the field direction), as shown in Fig. \ref{2}(b) in the main text. We introduce the dimensionless parameter $\beta$  defined as
 \begin{equation}
 \label{beta}
 \beta=1-\frac{R_{NL}(\alpha=90^{\circ})}{R_{NL}(\alpha=0^{\circ})}\; .
 \end{equation}
 In the limit $\gamma\ll 1$ we obtain up to lowest order in $\gamma$
 \begin{equation}
 \beta\approx\frac{\gamma_r}{2}\frac{L+\lambda}{\lambda}=\frac{\lambda(L+\lambda)\rho G_r}{t}\; , 
 \end{equation}
 while the lowest correction in $\gamma_i$ is of second order and therefore negligible in this approach. If we insert here the value for $G_r\approx4\times10^{11}\text{ }\Omega^{-1}m^{-2}$, we obtain an amplitude of the effect $\beta\approx 9.3\%$ ($\beta\approx 4.8\%$) for the LSV with $L=1600$ nm ($L=570$ nm), which is in good agreement with the experimentally obtained ones.  
  
 \begin{figure}[h]
\centering
\includegraphics[scale=1]{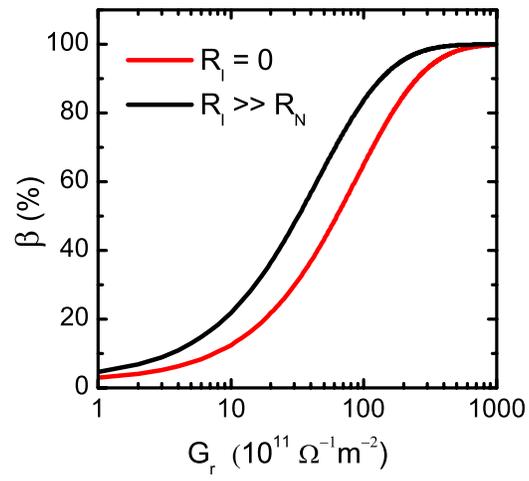}
\caption{\small{$\beta$ factor as a function of $G_r$ for the $R_I=0$ and $R_I\gg R_N$ cases.}}\label{S7}
\end{figure}

As explained above, all the previous results have been obtained assuming the high $R_I$ limit ($R_I\gg R_N$). However, if one allows for an arbitrary value of $R_I/R_N$, one should take into account  the possible  back flow of spin current in Eqs. (\ref{bc_inj}-\ref{bc_det}). An expression for $R_{NL}$ with arbitrary $R_I$ in a perpendicular magnetic field (without FMI) has been presented in Ref. \cite{s20}. In such a case only the lengths $\lambda$ and $\lambda_m$ enter in Eq. (\ref{1Deq}).  After inspection of the latter equation, it turns out that the general expression for $R_{NL}$ derived in Ref. \cite{s20} is also valid in the presence of the FMI layer, if one substitutes the magnetic length by $\lambda_1$ of Eq. (\ref{lambda1}). This result can be used to determine the parameter $\beta$ using Eq. (\ref{beta}). In Fig. \ref{S7} we show the dependence of $\beta$ as a function of $G_r$ in both the $R_I\gg R_N$ and $R_I=0$ cases. We see that for the value obtained from our measurements ($\beta\approx8\%$) $G_r$ is slightly larger (by a factor of $\sim 2$) in the transparent case. We can conclude from this that the actual value of $G_r$ lies between $4\times10^{11}\text{ }\Omega^{-1}m^{-2}$ and $8\times10^{11}\text{ }\Omega^{-1}m^{-2}$. 

It is also worth noticing that, according to Fig. \ref{S7}, in the hypothetical case that $G_r $ is of  the order of $10^{13}-10^{14}\text{ } \Omega^{-1}m^{-2}$ a full modulation ($\beta=100\%$) can be achieved. This means that  $R_{NL}$ can be switched between a finite value and $0$ by switching the field from $\alpha=0^{\circ}$ to $\alpha=90^{\circ}$, respectively.

\end{widetext}


\begin{thebibliography}{99}
\bibitem{36} The YIG was grown by liquid phase epitaxy on a (111) gadolinium gallium garnet (GGG) single crystal at Innovent e. V. (Jena, Germany).
\bibitem{39} A modulation in $R_NL$ originating from the rotation of magnetization of the Co electrodes (angles $\theta_{1}$, $\theta_{2}$ from the easy axis defined by the shape anisotropy) would be given by $R_{NL}\propto\cos\theta$, where $\theta$ is the relative angle between Co magnetizations. The observed $\sim$8\% modulation would correspond to $\theta\sim23^{\circ}$. Since $\theta=\theta_{1} - \theta_{2}$ in the P case, whereas in the AP case $\theta=180^{\circ} - \theta_{1} - \theta_{2}$, this would lead to a different amplitude modulation not observed in our experiment. Only in the special case where the narrower electrode does not rotate at all ($\theta_{2}\sim0$) would the modulations corresponding to the P and AP cases be identical. The wider electrode should then rotate as much as  $\theta_{1}\sim23^{\circ}$, which has been ruled out by MOKE microscopy as described in Ref. \cite{37}.
\end{thebibliography}
\end{document}